\documentstyle[12pt,epsf]{article}

\catcode`\@=11

\def\section{\@startsection{section}{1}{\z@}
  {3.5ex plus 1.0ex minus 0.2ex}{2.3ex plus .2ex}{\normalsize}}
\def\subsection{\@startsection{subsection}{2}{\z@}
  {3.25ex plus 1.0ex minus 0.2ex}{1.5ex plus 0.2ex}{\normalsize\bf}}
\def\subsubsection{\@startsection{subsubsection}{3}{\z@}
  {3.25ex plus 1.0ex minus 0.2ex}{1.5ex plus 0.2ex}{\normalsize\bf}}

\catcode`\@=12

\newcommand{\skipline}{\vspace{\baselineskip}}

\def\fun#1#2{\lower3.6pt\vbox{\baselineskip0pt\lineskip.9pt
  \ialign{$\mathsurround=0pt#1\hfil##\hfil$\crcr#2\crcr\sim\crcr}}}

\begin{document}

\begin{titlepage}

\begin{flushright}
IBM-HET-95-2
\end{flushright}

\skipline
\skipline
\skipline

\begin{center}
NUMERICAL EVIDENCE FOR THE OBSERVATION OF A SCALAR GLUEBALL
\end{center}

\skipline
\skipline

\begin{center}

J.\ Sexton\footnote{permanent address: Department of Mathematics,
Trinity College, Dublin 2,
Republic of Ireland},
A.\ Vaccarino\footnote{present address: Piazza Giovanetti 1,
Novara, 28100 Italy},
and D.\ Weingarten \\

IBM Research \\
P.O. Box 218, Yorktown Heights, NY 10598
\end{center}

\skipline
\skipline

\begin{center}
ABSTRACT
\end{center}

\begin{quotation}
We compute from lattice QCD in the valence (quenched) approximation the
partial decay widths of the lightest scalar glueball to pairs of
pseudoscalar quark-antiquark states.  These predictions and values
obtained earlier for the scalar glueball's mass are in good agreement
with the observed properties of $f_J(1710)$ and inconsistent with all
other observed meson resonances.
\end{quotation}

\skipline
\skipline

\end{titlepage}

It is generally believed that QCD predicts the existence of glueballs,
resonances composed mainly of chromoelectric field without a valence
quark-antiquark pair, occurring either as physical particles by
themselves or in linear combination with states which do include a
valence quark and antiquark.  Whether such states have been identified
so far in experiment remains ambiguous.  A crucial problem is that the
properties of glueballs are not expected to be drastically different
from the properties of flavor singlet bosons including valence quarks
and antiquarks.  Thus the identification in experiment of states with
large glueball contributions is difficult if not impossible in the
absence of a reliable evaluation of the properties predicted for
glueballs by QCD.  We believe the lattice formulation of QCD provides
the most reliable method now available for determining QCD's predictions
for the masses and decay couplings of hadrons.

Some time ago we reported \cite{Vaccarino} a value of 1740(71) MeV for
the valence (quenched) approximation to the infinite volume continuum
limit of lattice QCD predictions for the mass of the lightest scalar
glueball.  This result was obtained using ensembles of 25000 to 30000
gauge configurations on each of several different lattices.  An earlier
independent valence approximation calculation \cite{Livertal}, when
extrapolated to the continuum limit \cite{Weingarten94} following
Ref.~\cite{Vaccarino}, yields 1625(94) MeV for the lightest scalar
glueball mass.  This calculation used several different lattices
with ensembles of between 1000 and 3000 configurations each.  If the two
mass evaluations are combined, taking into account the correlations
between their statistical uncertainties arising from a common procedure
for converting lattice quantities into physical units, the result is
1707(64) MeV for the scalar glueball mass. Both the mass prediction with
larger statistical weight and the combined mass prediction are in good
agreement with the mass of $f_J(1710)$ and are strongly inconsistent
with all but $f_0(1500)$~\cite{Amsler1} among the established flavor
singlet scalar resonances. For $f_0(1500)$ the disagreement is still by
more than three standard deviations.

The valence approximation, used in the mass calculation of
Refs.~\cite{Vaccarino} \cite{Livertal}, may be viewed as replacing the
momentum and frequency dependent color dielectric constant arising from
quark-antiquark vacuum polarization with its zero-momentum,
zero-frequency limit \cite{Weingarten82}. This approximation is expected
to be fairly reliable for long-distance properties of hadrons. For
example, the infinite volume continuum limits of the valence
approximation to the masses of eight low-lying hadrons composed of quarks
and antiquarks differ from experiment by amounts ranging up to 6\%
\cite{Butler}.  A 6\% error in the glueball mass would be 100 MeV and,
according to an adaptation of an argument giving a negative sign for the
valence approximation error in $f_{\pi}$~\cite{Butler}, the sign of this
error is also expected to be negative. Thus the scalar glueball in full
QCD should lie above the valence approximation mass, and correcting the
error in the valence approximation should not drastically change the
comparison with experiment.

The most likely interpretation of $f_0(1500)$, we believe, is not as a
glueball~\cite{Amsler2} but as a state composed largely of an
$s\overline{s}$ quark-antiquark pair.  The $s\overline{u}$ scalar and
tensor are nearly degenerate at about 1430 MeV. Thus the $s\overline{s}$
scalar and tensor should lie close to each other somewhere above 1430
MeV. Since the $s\overline{s}$ tensor has been identified at 1525 MeV,
an $s\overline{s}$ scalar at 1500 MeV would be quite natural.

The crucial question not answered by the mass results, however, is
whether the decay width of the lightest scalar glueball is small enough
for this particle actually to be identified in experiment. In addition,
it is sometimes argued that since glueballs are flavor singlets they
should have the same couplings to $2 \pi_0$, to $2 K_L$, and to $2
\eta$. This expectation is violated by $f_J(1710)$ decay
couplings.

\begin{figure}
\begin{center}
\leavevmode
\epsfxsize=\textwidth
\epsfbox{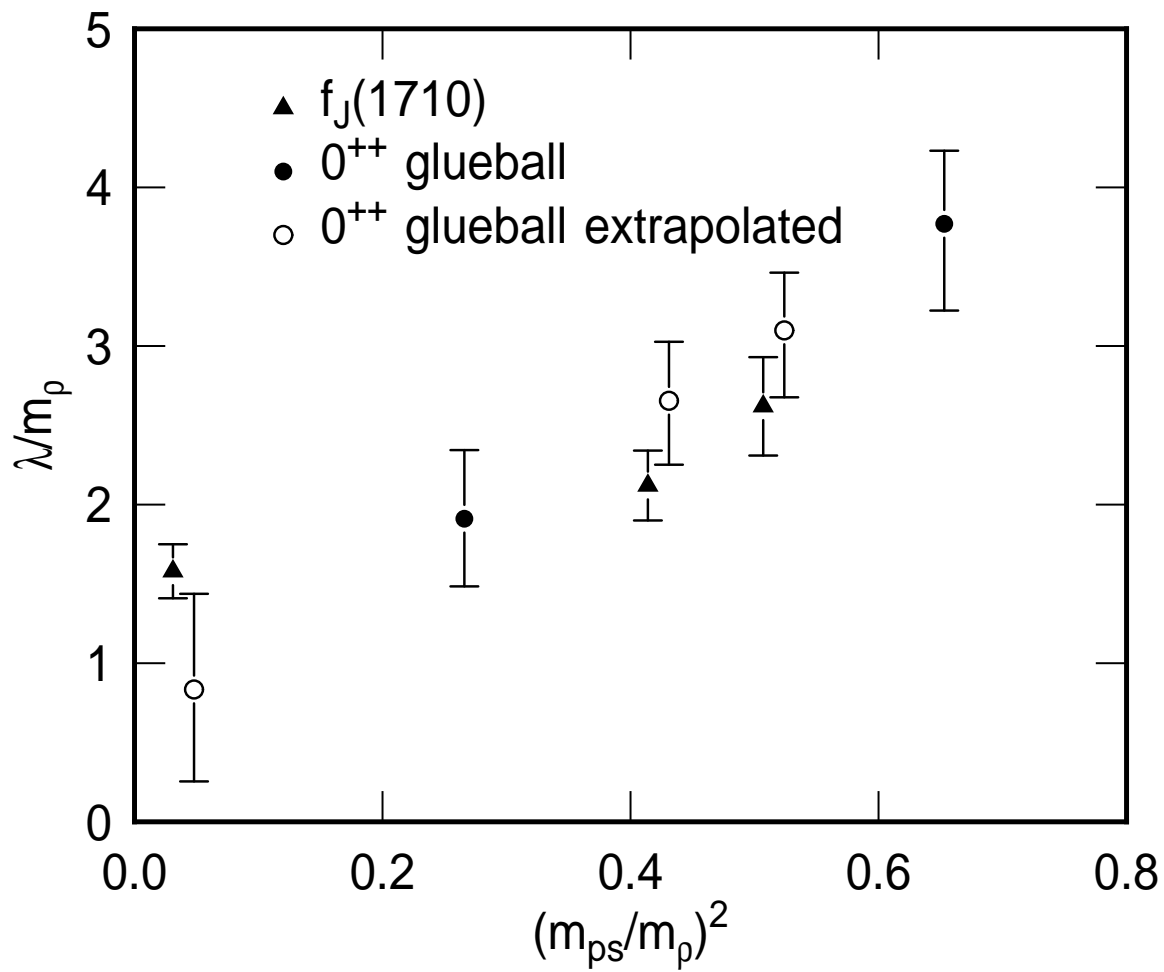}
\end{center}
\caption{
Decay couplings.}
\label{fig:lambdas}
\end{figure}

In the present article we report the first lattice QCD calculation of
the valence (quenched) approximation to the partial decay widths of the
lightest scalar glueball to pairs of pseudoscalar quark-antiquark
states. The calculation is done with 10500 gauge configurations on a
single lattice, $16^3 \times 24$, at $\beta = 5.70$ corresponding to
inverse lattice spacing $a^{-1} = 1.35$ GeV. We believe this lattice has
spacing sufficiently small and volume sufficiently large to give partial
widths within 30\% of their infinite volume continuum limits.  The
predicted decay couplings, combined with the mass prediction of 1740(71)
MeV, give a total two-pseudoscalar decay width of 108(29) MeV for the
scalar glueball.  With any reasonable guess concerning the scalar
glueball's branching fraction to multibody decay modes, the resulting
total decay width is well below 200 MeV and therefore small enough for
the scalar glueball to be identified in experiment. In fact, the
predicted total two-pseudoscalar decay width, and individual couplings
to $2 \pi_0$, to $2 K_L$, and to $2 \eta$ are all in good agreement with
properties of $f_J(1710)$ and inconsistent with all other established
flavor singlet scalar resonances. A comparison of our results with data
for $f_J(1710)$ \cite{Lind} is shown in Figure~\ref{fig:lambdas}.

Glueballs found in the valence approximation, according to one simple
interpretation, contain no admixture of configurations with valence
quarks or antiquarks.  Thus we consider the agreement between the mass
and decay couplings found in the valence approximation and the observed
mass and decay couplings of $f_J(1710)$ to be strong evidence that this
state is largely a scalar glueball with at most some relatively
smaller amplitude for configurations including valence quark-antiquark
pairs.

The calculations presented here were carried out on the GF11 parallel
computer \cite{Weingarten90} at IBM Research and took approximately two
years to complete at a sustained computation rate of between 6 and 7
Gflops. A preliminary version of this work is discussed in
Ref.~\cite{Sexton95}.

In the remainder of this paper we describe our method for determining
scalar glueball decay couplings then present our numerical results.

To evaluate glueball decay couplings we work with a euclidean lattice
gauge theory, on a lattice $L^3 \times T$, with the plaquette action for
the gauge field, and the Wilson action for quarks. It is convenient
initially to assume exact flavor SU(3) symmetry for the quark mass
matrix.  With each gauge configuration fixed to lattice Coulomb gauge,
we construct a collection of smeared fields.  We describe smearing only
for the particular choice of parameters actually used in the decay
evaluation.  Let $U_i(x)$ for a space direction $i = 1, 2, 3,$ be a
smeared link field~\cite{Vaccarino} given by the average of the 9 links
in direction $i$ from the sites of the (3 site) x (3 site) square
oriented in the two positive space directions orthogonal to $i$ starting
at site $x$. Let $V_{ij}(x)$ be the trace of the product around the
outside of a (3 link) x (3 link) square $tr[ U_i(x) U_j(x + 3\hat{i} +
2\hat{j}) U^{\dagger}_i(x + 5\hat{j}) U^{\dagger}_i(x - 2\hat{i} +
2\hat{j})]$, where $\hat{i}$ is an $i$-direction unit vector. Define the
zero-momentum scalar glueball operator $g(t)$ to be the sum of the
$V_{ij}(x)$ for all $i, j$ and $x$ with time component $t$.  Let the
quark and antiquark fields $\overline{\Psi}(x)$ and $\Psi(x)$ be Wilson
quark and antiquark fields smeared~\cite{Butler} by convoluting the
local Wilson fields with a space direction gaussian, invariant under
lattice rotations and with mean-square radius 6.0. The smeared
pseudoscalar field $\pi_i(x)$ with flavor index $i$ is
$\overline{\Psi}(x) \gamma^5 \Lambda_i \Psi(x)$, where $\Lambda_i$ is a
Gell-Mann flavor matrix.  Let $\tilde{\pi}_i(\vec{k},t)$ be the Fourier
transform of $\pi_i(x)$ on the time $t$ lattice hyperplane.

Define $E^{\pi}_1$ and $E^{\pi}_2$ to be the energy of a single
pseudoscalar at rest or with momentum magnitude $|\vec{k}| = 2 \pi/ L$,
respectively. The field strength renormalization constant $\eta^{\pi}_1$
is defined by the requirement that for large $t$ the vacuum expectation
value $< \tilde{\pi}^{\dagger}_i( 0,t) \tilde{\pi}_i( 0,0)>$ approaches
$(\eta^{\pi}_1)^2 L^3 exp[ -E_1 t]$. Define $\eta^{\pi}_2$ similarly
from a pseudoscalar field with momentum magnitude $|\vec{k}| = 2 \pi/
L$.  In the valence approximation, the glueball is stable so that its
mass $E^g$ and field strength renormalization constant $\eta^g$ can be
defined by the requirement that, for large $t$, $<g(t) g(0))>$
approaches $(\eta^g)^2 L^3 exp( -E^g t)$.

{}From pseudoscalar fields at position 0 and times $t_i$, define the
two-pseudoscalar, flavor singlet field $\Pi( t_1, t_2)$ to be $(16)^{-1/2}
\sum_i \pi_i(0,t_1) \pi_i(0,t_2)$, where the sum over $i$ runs from 1 to 8.
Let the zero-momentum, two-pseudoscalar flavor singlet field
$\tilde{\Pi}_1( t_1, t_2)$ be $(16)^{-1/2} \sum_i \tilde{\pi}_i(0,t_1)
\tilde{\pi}_i(0,t_2)$. Define the two-pseudoscalar field
$\tilde{\Pi}_2( t_1, t_2)$ to be $(24)^{-1/2} \sum_{i \vec{k}}
\tilde{\pi}_i(\vec{k},t_1) \tilde{\pi}_i(-\vec{k},t_2)$ where the sum
for $\vec{k}$ is over the three positive orientations with $|\vec{k}| =
2 \pi/ L$.

Let $| 1>$ and $| 2>$ be, respectively, the lowest and second lowest
energy flavor singlet, rotationally invariant two-pseudoscalar states.
Both states are normalized to 1.  Let $E^{\pi\pi}_i$ be the energy of $|
i>$.  Define the amplitudes $\eta^{\pi \pi}_{ij}(t)$ to be $L^{-3} < i|
\tilde{\Pi}_j( t, 0) | \Omega >$. For large $t$, $\eta^{\pi
\pi}_{ij}(t)$ has the asymptotic form $\eta^{\pi \pi}_{ij}
exp(-E^{\pi}_j t)$.  The diagonal coefficients $\eta^{\pi \pi}_{11}$ and
$\eta^{\pi \pi}_{22}$ are expected to be larger than the off-diagonal
$\eta^{\pi \pi}_{21}$ and $\eta^{\pi \pi}_{12}$, respectively. As a
consequence of the interaction between pairs of pseudoscalars, however,
the off-diagonal coefficients will not be zero.

Connected three-point functions from which coupling constants can be
extracted are now given by $T_i( t_g, t_{\pi})$ defined as $< g(t_g)
\tilde{\Pi}_i( t_{\pi}, 0) > - < g(t_g)> < \tilde{\Pi}_i( t_{\pi}, 0)
>$.  If the quark mass, and thus the pseudoscalar mass, is chosen so
that $E^{\pi\pi}_1$ is equal to $E^g$, the lightest intermediate state
which can appear between the glueball and pseudoscalars in a transfer
matrix expression for $T_1( t_g, t_{\pi})$ is $| 1>$. Thus for large
enough $t_g$ with $t_{\pi}$ fixed, $T_1( t_g, t_{\pi})$ will be
proportional to the coupling constant of a glueball to two pseudoscalars
at rest.  If the quark mass is chosen so that $E^{\pi\pi}_2$ is equal to
$E^g$, however, the lightest intermediate state which can appear between
the glueball and pseudoscalars in a transfer matrix expression for $T_2(
t_g, t_{\pi})$ we still expect to be $| 1>$, not $| 2>$, since
$\eta^{\pi \pi}_{1 2}(t)$ is expected not to be zero.  To obtain from
$T_2( t_g, t_{\pi})$ the coupling of a glueball to two pseudoscalars
with momenta of magnitude $2 \pi L^{-1}$, the contribution to $T_2( t_g,
t_{\pi})$ arising from the $| 1>$ intermediate state must be removed.

{}From the three-point functions we therefore define the amplitudes
\begin{eqnarray}
\label{defs}
S_i( t_g, t_{\pi})  = T_i( t_g, t_{\pi})  -
\frac{\eta^{\pi \pi}_{ji}(t_{\pi})}
{\eta^{\pi \pi}_{jj}(t_{\pi})} T_j( t_g, t_{\pi}),
\end{eqnarray}
for $(i,j)$ of either (1,2) or (2,1).  In $S_2( t_g, t_{\pi})$ the
contribution of the undesirable $| 1>$ intermediate state has been
canceled.  In $S_1( t_g, t_{\pi})$ a contribution from the intermediate
state $| 2>$ has been canceled.  Although the subtraction in $S_1( t_g,
t_{\pi})$ is irrelevant for large enough $t_g$, we expect that as a
result of this subtraction $S_1( t_g,t_{\pi})$ will approach its large
$t_g$ behavior more rapidly than does $T_1( t_g, t_{\pi})$.

An additional intermediate state which can also appear in a transfer
matrix expression for either $T_i( t_g, t_{\pi})$ is the isosinglet
scalar bound state of a quark and an antiquark. For the parameter values
used in the present calculation we have found that this state has a mass
in lattice units above 1.25 while the scalar glueball mass is 0.972(44).
Thus for large enough $t_g$ the scalar quark-antiquark state will make
only its appropriate virtual contribution and does not require an
additional correction.

At large $t_g$ and $t_{\pi}$, the three-point functions become
\begin{eqnarray}
\label{Sasym}
S_i( t_g, t_{\pi}) \rightarrow
\frac{c_i \sqrt{3} \lambda_i \eta^g \eta^{\pi \pi}_{ii}
(1 - r) L^3} {\sqrt{8 E^g (E^{\pi}_i)^2}}
s_i(t_g, t_{\pi}),
\end{eqnarray}
where $c_1 = 1/\sqrt{2}$, $c_2 = \sqrt{3}$, $r$ is $(\eta^{\pi \pi}_{12}
\eta^{\pi \pi}_{21}) / (\eta^{\pi \pi}_{11} \eta^{\pi \pi}_{22})$ and
$\lambda_1$ and $\lambda_2$ are the glueball coupling constants to a
pair of pseudoscalars at rest or with momenta of magnitude $2 \pi
L^{-1}$, respectively.  The factors $\eta^{\pi \pi}_{ij}$ are given by
the large $t$ behavior of $\eta^{\pi \pi}_{ij}(t)$ as discussed earlier.
For $T \gg t_g \ge t_{\pi}$, the factors $s_i(t_g, t_{\pi})$ are
\begin{eqnarray}
\label{defsi}
s_i(t_g, t_{\pi}) & = &
\sum_t exp[ -E^g |t - t_g| - E^{\pi}_i |t| - \nonumber \\
& & E^{\pi}_i |t - t_{\pi}| - \delta_i( t, t_{\pi})|t - t_{\pi}|],
\end{eqnarray}
where, for $t \ge t_{\pi}$, $\delta_i(t, t_{\pi})$ is the binding energy
$E^{\pi \pi}_i - 2 E^{\pi}_i$ and otherwise $\delta_i(t, t_{\pi})$ is 0.

The coupling constants in Eq.~(\ref{Sasym}) have been identified by
comparing $S_i(t_g, t_{\pi})$ with the three-point functions arising
from a simple phenomenological interaction lagrangian. This procedure is
correct to leading order in the coupling constants. A similar relation
used to find coupling constants among hadrons containing quarks has
recently yielded several predictions in good agreement with
experiment~\cite{decays}.  The $\lambda_i$ are normalized so that in the
continuum limit they become, up to a factor of $-i$, Lorentz-invariant
decay amplitudes with the standard normalization convention used in the
section on kinematics of the Review of Particle Properties.

To obtain values of $\lambda_i$ from Eq.~(\ref{Sasym}) we need the
amplitudes $\eta^{\pi \pi}_{ij}(t)$.  These we determine from
propagators for two-pseudoscalar states.  Define two-pseudoscalar
propagators $C_i(t_1,t_2)$ to be $< \Pi(t_1 + 2 t_2, t_1 + t_2)
\tilde{\Pi}_i( t_2, 0)>$.  For moderately large values of $t_1$, these
amplitudes approach
\begin{eqnarray}
\label{Casym}
C_i(t_1,t_2) & = &
C_{1i} exp( -E^{\pi\pi}_1 t_1) + C_{2i} exp( -E^{\pi\pi}_2 t_1), \\
\label{Cxasym}
C_{ij} & = &
\eta^{\pi \pi}_{i1}(t_2) \eta^{\pi \pi}_{ij}(t_2) + \sqrt{6}
\eta^{\pi \pi}_{i2}(t_2) \eta^{\pi \pi}_{ij}(t_2).
\end{eqnarray}
{}From these expressions the required $\eta^{\pi \pi}_{ij}(t)$ can be
extracted.

The $\eta^{\pi \pi}_{ij}(t)$ in Eq.~(\ref{Sasym}) serve, among other
purposes, to correct for the interaction between the two pseudoscalars
produced by a glueball decay. In the valence approximation this
interaction does not include the production and annihilation of virtual
quark-antiquark pairs.  Correspondingly, in the numerical evaluation of
$C_i(t_1,t_2)$ from quark propagators, we include only terms in which
all initial quarks and antiquarks propagate through to some final quark
or antiquark. Terms in the two-pseudoscalar propagator in which initial
quarks propagate to initial antiquarks can be shown to correspond to
processes missing from glueball decay in the valence approximation. For
very large $t_1$ and $T$, the $C_i(t_1,t_2)$ are given by a sum of two
terms each of which is a slightly more complicated version of one of the
exponentials in Eq.~(\ref{Casym}).  This complication occurs, for
example, because in the valence approximation the exchange of a $\rho$
between the pseudoscalars produced in a glueball decay is not iterated
in the same way as in full QCD.  Each term in Eq.~(\ref{Casym}) holds
without modification if $| E^{\pi \pi}_i - 2 E^{\pi}_i |^2 t_1^2 / 2 \ll
< 1.$ The intervals of $t_1$ we use to determine the $\eta^{\pi
\pi}_{ij}$ fall well within this limitation.
In any case, as we will discuss below, the measured values of $\eta^{\pi
\pi}_{ij}$ turn out to be close to their values for noninteracting
pseudoscalars. As a consequence, the corrections due to interations
between the decay pseudoscalars which the $\eta^{\pi \pi}_{ij}$
contribute to the predicted values of $\lambda_i$ are comparatively
small.

We now turn to our numerical results.  At $\beta = 5.7$ on a $16^3
\times 24$ lattice, with an ensemble of 10500 independent configurations,
we determined glueball and single pseudoscalar energies and
renormalization constants following Refs.~\cite{Vaccarino} and
\cite{Butler}, respectively. For $E^g$, as mentioned above,
we found $0.972 \pm 0.044$.  On a lattice of size $16^3 \times 40$ we
then evaluated the two-pseudoscalar propagator $C_i( t_1, t_2)$ at
$\kappa = 0.1650$ using 100 independent configuration, and at $\kappa =
0.1675$ using 875 independent configurations. Fitting the $t_1$
dependence of $C_i(t_1,t_2)$ to Eqs.~(\ref{Casym}) and (\ref{Cxasym}),
we determined $E^{\pi \pi}_i$ and $\eta^{\pi \pi}_{ij}(t_2)$ for a range
of different $t_2$. At $\kappa = 0.1650$ we obtained results for $0 \le
t_2 \le 4$, and at 0.1675 we found results for $0 \le t_2 \le 5$.  The
values of $E^{\pi \pi}_i$ were statistically consistent with being
independent of $t_2$ in all cases. The $\eta^{\pi \pi}_{ij}(t_2)$ were
consistent with the asymptotic form $\eta^{\pi \pi}_{ij} exp( -E^{\pi}_j
t_2)$ in all cases for $t_2 \ge 2$. At $\kappa = 0.1650$ for $E^{\pi
\pi}_1$ we obtained 0.908(5), giving glueball decay to $| 1>$
nearly on mass shell. At $\kappa = 0.1675$ for $E^{\pi \pi}_2$ we found
$0.893^{+0.044}_{-0.004}$, giving glueball decay to $| 2>$ nearly on
mass shell.  For the normalized ratios $\hat{\eta}^{\pi \pi}_{ij}$
defined as $\eta^{\pi \pi}_{ij}/ (\eta^{\pi}_j)^2$, at $\kappa = 0.1650$
we obtained for $ij$ of 11, 12, 21 and 22, the values 0.988(30),
0.091(8), -0.087(8), and 1.065(13), respectively.  At $\kappa = 0.1675$
we found 1.050(21), 0.107(6), -0.112(8), 1.053(53).  For noninteracting
pseudoscalars $\hat{\eta}^{\pi \pi}_{ij}$ is 1 for $i = j$ and 0
otherwise.  Our data is close to these values. The final value of
$\lambda_1$ is changed by less than 1 standard deviation and the final
$\lambda_2$ is changed by less than 2 standard deviations if we ignore
the determination of $\hat{\eta}^{\pi \pi}_{ij}$ and simply use the the
noninteracting values.

\begin{figure}
\begin{center}
\leavevmode
\epsfxsize=\textwidth
\epsfbox{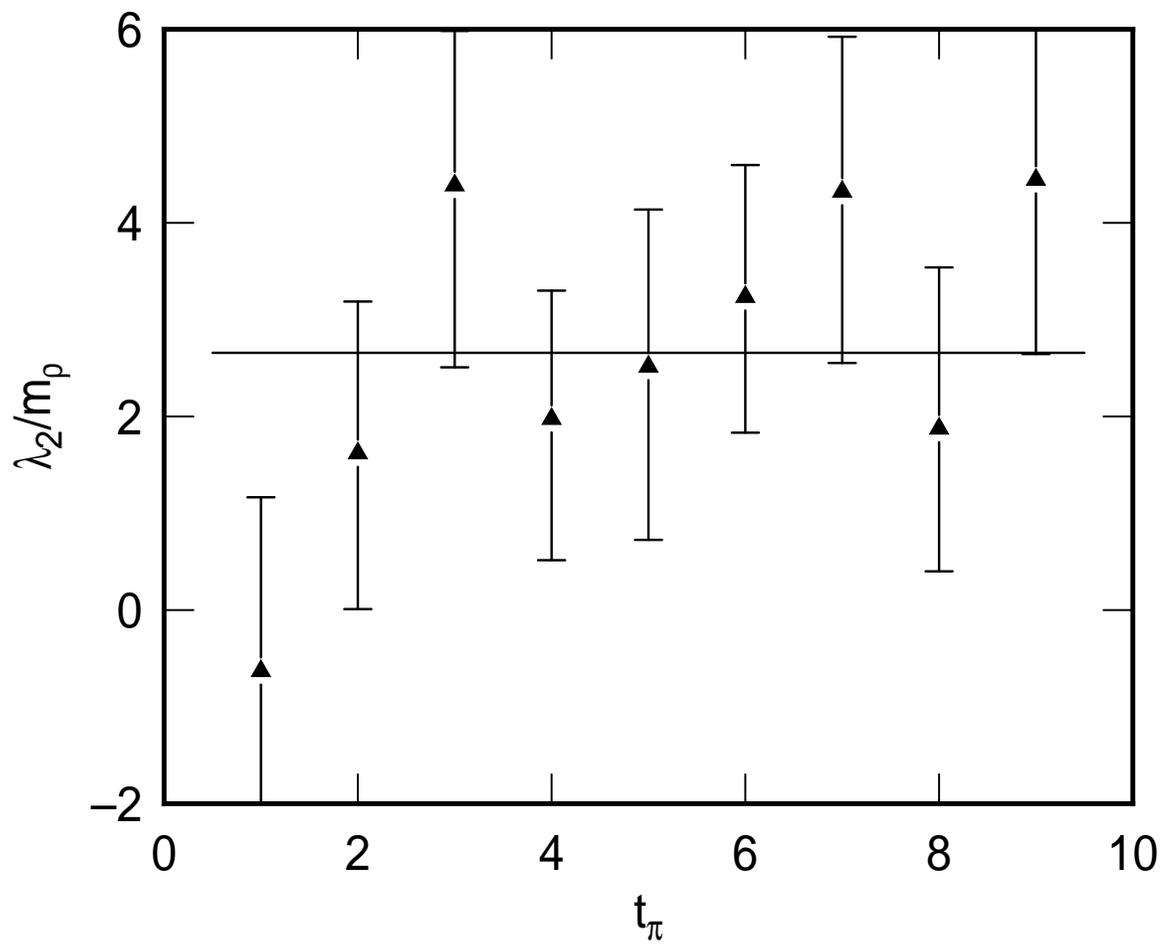}
\end{center}
\caption{
$\lambda_2$ for $t_g - t_{\pi} = 2$.
}
\label{fig:g1t2}
\end{figure}

\begin{figure}
\begin{center}
\leavevmode
\epsfxsize=\textwidth
\epsfbox{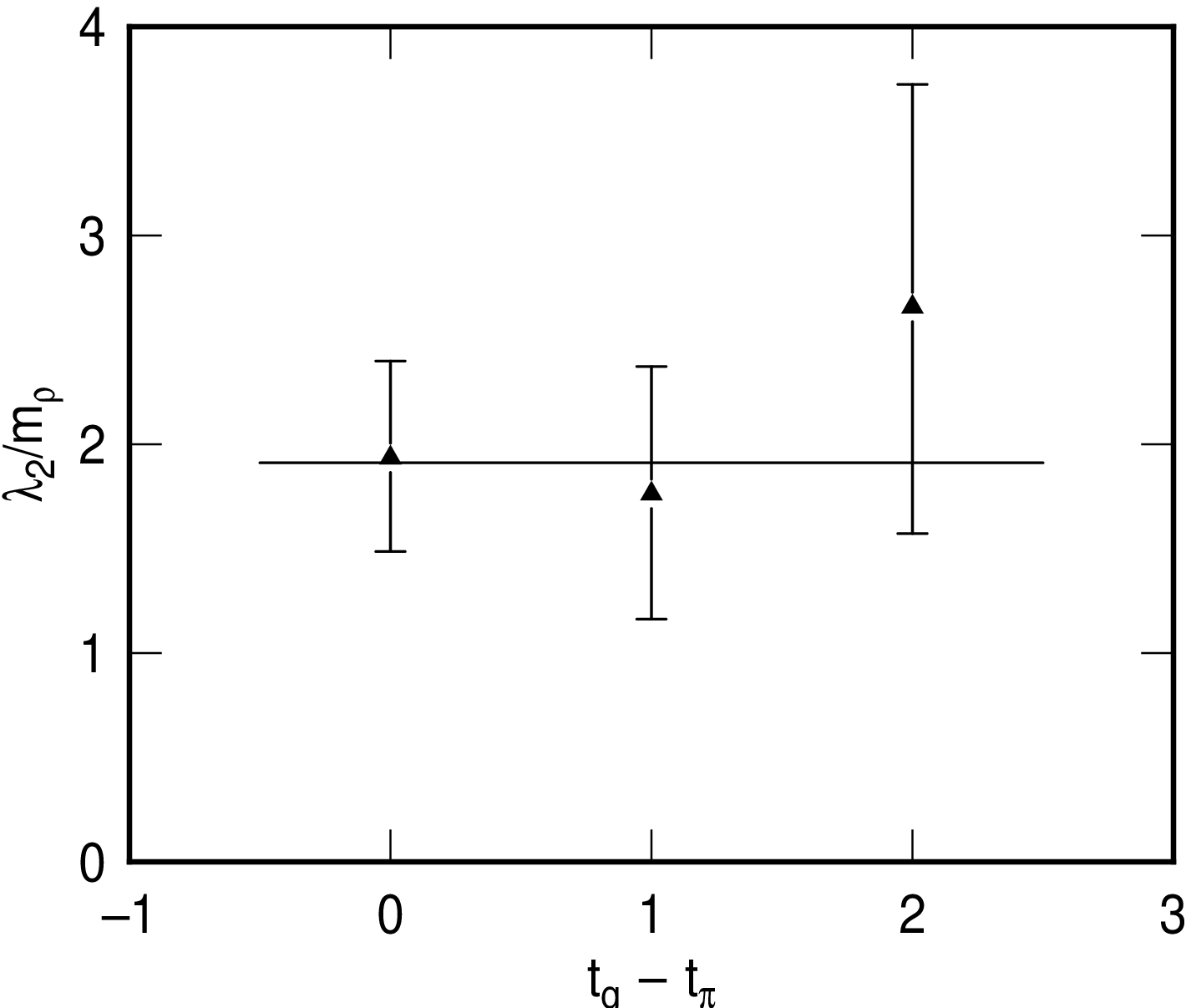}
\end{center}
\caption{
$\lambda_2$ fitted on $2 \le t_{\pi} \le 6$
as a function of $t_g - t_{\pi}$ for the fitting interval.
}
\label{fig:g1t012}
\end{figure}

{}From our 10500 configuration ensemble on a $16^3 \times 24$ lattice, we
evaluated $S_1$ and $S_2$ for glueball decay on mass shell at $\kappa$
of 0.1650 and 0.1675, respectively. We obtained statistically
significant results for $0 \le t_g - t_{\pi} \le 2$ with $0 \le t_{\pi}
\le 8$. At each point within this range we then determined
effective $\lambda_i$ using Eq.~(\ref{Sasym}).  We found $\lambda_1$ and
$\lambda_2$ statistically consistent with being constant for $t_{\pi}
\ge 3$ and $t_{\pi} \ge 2$, respectively, and all values of $t_g -
t_{\pi}$.  Figure~\ref{fig:g1t2}, for example, shows effective
$\lambda_2$ in units of the $\rho$ mass as a function of $t_{\pi}$ for
$t_g - t_{\pi} = 2$, in comparison to a fit with $2 \le t_{\pi} \le 6$,
$t_g - t_{\pi} = 2$.  Figure~\ref{fig:g1t012} shows fitted values of
$\lambda_2$ on the interval $2 \le t_{\pi} \le 6$ for fixed $t_g -
t_{\pi}$ of 0, 1 or 2.  To extract final values of $\lambda_i$, we tried
fits to all rectangular intervals of data including at least 4 values of
$t_{\pi}$ and at least 2 values of $t_g - t_{\pi}$. For each $\lambda_i$
we chose the fit giving the lowest value of $\chi^2$ per degree of
freedom. The window determined in this way for $\lambda_1$ is
$3 \le t_{\pi} \le 7$ with $1 \le t_g - t_{\pi} \le 2$, and for
$\lambda_2$ is $2 \le t_{\pi} \le 6$ with $0 \le t_g - t_{\pi}
\le 1$.  The horizontal line in Figure~\ref{fig:g1t012} shows the final
value of $\lambda_2$.  Over the full collection of windows we examined,
the fitted results varied from our final results by at most 1 standard
deviation. We believe our best fits provide reasonable estimates of the
asymptotic coefficients in Eq.~(\ref{Sasym}).

So far our discussion has been restricted to QCD with u, d and s quark
masses degenerate. An expansion to first order in the quark mass matrix
taken around some relatively heavy SU(3) symmetric point gives glueball
decay couplings for $\pi$'s, K's and $\eta$'s which are a common
linear function of each meson's average quark mass. Since meson masses
squared are also nearly a linear function of average quark mass, the
decay couplings are a linear function of meson masses squared. Thus from
a linear fit to our predictions for decay couplings as a function of
pseudoscalar mass squared at unphysical degenerate values of quark
masses we can extrapolate decay couplings for physical nondegenerate
values of quark masses.  From this linear fit a prediction can also be
made for the decay coupling of the scalar glueball to $\eta + \eta'$, if
we ignore the contribution to the decay from the process in which the
$\eta$ quark and antiquark are connected to each other by one propagator
and the $\eta'$ quark and antiquark are connected to each other by a
second propagator.

Figure~\ref{fig:lambdas} shows predicted coupling constants as a fuction
of predicted meson mass squared along with linear extrapolations of the
predicted values to the physical $\pi$, K and $\eta$ masses, in
comparison to observed decay couplings\cite{Lind} for decays of
$f_J(1710)$ to pairs of $\pi$'s, K's and $\eta$'s.  Masses and decay
constants are shown in units of the $\rho$ mass. Our predicted width for
the scalar glueball decay to $\eta + \eta'$ is 6(3) MeV. For the ratio
$\lambda_{\eta \eta'} / \lambda_{\eta \eta}$ we get 0.52(13). We predict
a total width for glueball decay to pseudoscalar pairs of 108(28) MeV,
in comparison to 99(15) MeV for $f_J(1710)$.


\begin{thebibliography}{9}

\bibitem[*]{Trinity} Permanent address:
Dept. of Mathematics, Trinity College, Dublin 2, Ireland.
\bibitem{Vaccarino} H.\ Chen, J.\ Sexton, A.\ Vaccarino
and D.\ Weingarten, Nucl.\ Phys.\ B (Proc.\ Suppl.) 34, 357 (1994).
\bibitem{Livertal} G.\ Bali, K.\ Schilling, A.\ Hulsebos,
A.\ Irving, C.\ Michael, and P.\ Stephenson,
Phys.\ Lett.\ B 309, 378 (1993).
\bibitem{Weingarten94} D.\ Weingarten, Nucl.\ Phys.\ B (Proc.\ Suppl.)
34, 29 (1994).
\bibitem{Amsler1} C.\ Amsler, et al., Phys.\ Lett.\ B355, 425 (1995).
\bibitem{Weingarten82} D.\ Weingarten, Phys.\ Lett.\ 109B,
57 (1982).
\bibitem{Butler} F.\ Butler, H.\ Chen, J.\ Sexton, A.\ Vaccarino, and
D.\ Weingarten, Nucl.\ Phys.\ B 430, 179 (1994); Nucl.\ Phys.\ B 421,
217 (1994).
\bibitem{Amsler2} C.\ Amsler and F.\ Close, Phys.\ Lett.\ B353, 385
(1995).
\bibitem{Lind} S.\ Lindenbaum and R.\ Longacre,
Phys.\ Letts.\ B274, 494 (1992).
\bibitem{Weingarten90} D.\ Weingarten, Nucl.\ Phys.\ B (Proc.\ Suppl.) 17,
272 (1990).
\bibitem{Sexton95} J.\ Sexton, A.\ Vaccarino and D.\ Weingarten,
Nucl.\ Phys.\ B (Proc.\ Suppl.) 42, 279 (1995).
\bibitem{decays} S.\ Gottlieb, et al., Phys.\ Lett.\ 134B, 346 (1984);
R.\ Altmeyer, et al., Nucl.\ Phys.\ B (Proc.\ Suppl.)  34, 373 (1994);
K.\ Liu, et al.\, Phys.\ Rev. Lett.\ 74, 2172 (1995).

\end{thebibliography}
\end{document}